\documentclass[journal]{IEEEtran}
\usepackage{amsmath,amsfonts}
\usepackage{algorithmic}
\usepackage{algorithm}
\usepackage{array}
\usepackage{textcomp}
\usepackage{stfloats}
\usepackage{url}
\usepackage{verbatim}
\usepackage{graphicx}
\usepackage{cite}
\usepackage{booktabs} % for \toprule, \midrule, \bottomrule
\usepackage{tabularx} % for tabularx environment
\usepackage{multirow}
\usepackage{xcolor}
\usepackage[bookmarks=false,
colorlinks=true,     % Colours links instead of boxes
linkcolor=blue,      % Hyperlink color for links
citecolor=blue,      % Hyperlink color for citations
urlcolor=black       % Hyperlink color for URLs
]{hyperref}

\usepackage{orcidlink}
\makeatletter
\def\ps@IEEEtitlepagestyle{%
  \def\@oddfoot{%
    \hbox to \textwidth{%
      \hfil
      \parbox{\textwidth}{\centering\scriptsize
      \textcopyright~2025 IEEE. Personal use of this material is permitted. Permission from IEEE must be obtained for all other uses, in any current or future media, including reprinting/republishing this material for advertising or promotional purposes, creating new collective works, for resale or redistribution to servers or lists, or reuse of any copyrighted component of this work in other works. DOI: \href{https://doi.org/10.1109/TVLSI.2025.3571677}{10.1109/TVLSI.2025.3571677}}%
      \hfil
    }%
  }%
  \def\@evenfoot{}%
}
\makeatother

\begin{document}

\title{A Scalable FPGA Architecture With Adaptive Memory Utilization for GEMM-Based Operations}

\author{Anastasios Petropoulos\,\orcidlink{0000-0003-1669-5233} and Theodore Antonakopoulos\,\orcidlink{0000-0002-7863-1051}%
\thanks{This work was supported by the European Union (Horizon Europe) Project NeuroSoC under Grant 101070634. \textit{(Corresponding author: \mbox{Anastasios} Petropoulos.)}}%
\thanks{The authors are with the Department of Electrical and Computer Engineering, University of Patras, 26504 Rio-Patras, Greece (e-mail: a.petropoulos@ece.upatras.gr; antonako@upatras.gr).}%
\thanks{}}

\markboth{}%
{Petropoulos \MakeLowercase{\textit{et al.}}: A Scalable FPGA Architecture With Adaptive Memory Utilization for GEMM-Based Operations}

% \IEEEpubid{1063-8210 © 2024 IEEE. Personal use is permitted, but republication/redistribution requires IEEE permission.
%  See https://www.ieee.org/publications/rights/index.html for more information.}
\maketitle

\begin{abstract}
Deep neural network (DNN) inference relies increasingly on specialized hardware for high computational efficiency. This work introduces a field-programmable gate array (FPGA)-based dynamically configurable accelerator featuring systolic arrays, high-bandwidth memory, and UltraRAMs. We present two processing unit (PU) configurations with different computing capabilities using the same interfaces and peripheral blocks. By instantiating multiple PUs and employing a heuristic weight transfer schedule, the architecture achieves notable throughput efficiency over prior works. Moreover, we outline how the architecture can be extended to emulate analog in-memory computing (AIMC) devices to aid next-generation heterogeneous AIMC chip designs and investigate device-level noise behavior. Overall, this brief presents a versatile DNN inference acceleration architecture adaptable to various models and future FPGA designs.
\end{abstract}


% \IEEEpeerreviewmaketitle

\vspace{-0.25cm}
\section{Introduction}
\vspace{-0.05cm}
\IEEEPARstart{D}{eep} neural network (DNN) inference demanded the design of specialized hardware platforms, with numerous architectures optimized for high throughput and/or low latency \cite{yu2019opu, abdelfattah2018dla, xing2019dnnvm, d2022xdnn, Wu2024Amoeba}. Field-programmable gate arrays (FPGAs), due to their reconfigurable nature, enable domain-specific accelerators that can be adapted to the requirements of different network architectures. Many FPGA-based designs employ systolic arrays (SAs) for core multiply-accumulate operations, whether implemented via high-level synthesis \cite{wang2021autosa} or register-transfer level (RTL) approaches \cite{li2024revealing}. Despite notable performance achievements, some solutions overlook the physical layout of the FPGA resources and their interconnect structure, resulting in limited performance and non-optimal clock frequencies. Others utilize architectural features such as cascade paths, DSP packing strategies, and clock-domains separation techniques to harness more of the FPGA capabilities, delivering higher performance \cite{samajdar2019scaling, d2022xdnn, li2024revealing}. However, on-chip memory constraints, bandwidth limitations, and variations in resources availability across FPGAs can restrict portability and design reusability.

In this work, we propose a highly adaptable systolic-array-based processing unit (PU), which leverages high-bandwidth memory (HBM) for high sustained data transfer rates and UltraRAM (URAM) for large-scale weights storage. The PU is parameterized for versatility, supporting diverse FPGA devices with URAM and HBM resources. For improved resources utilization, we implemented two PU configurations—one with high DSP usage and another with half of this capacity—using the same interfaces and peripheral blocks. Also, we integrated multiple PUs on an Alveo FPGA, achieving higher throughput and energy efficiency on ResNet models over prior works. Finally, we devise how the architecture can be extended to emulate analog in-memory computing (AIMC) devices.

% columnwidth
\begin{figure}[!t]
\centering
\includegraphics[width=1.0\columnwidth]{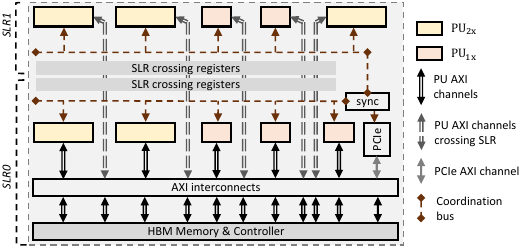}
\vspace{-0.7cm}
\caption{The system architecture of multiple PUs on an Alveo U50 FPGA.}
\label{fig:arch_overview}
\vspace{-0.5cm}
\end{figure}

% \IEEEpubidadjcol

\vspace{-0.2cm}
\section{Hardware Architecture} \label{sec:hw_arch}
\vspace{-0.05cm}
Our architecture prioritizes scalability and computational efficiency to support convolutional (Conv) and fully connected (FC) layers within the same PU. This design is intended to accommodate evolving DNN workloads for high-throughput computations rather than minimizing metrics such as energy consumption. To achieve that, fast off-chip HBM is crucial to sustain a high-throughput dataflow within a single PU and/or among multiple PUs and to support rapid weight updates to on-chip memories. The system architecture with multiple PUs in two configurations, along with their AXI channel interfaces to the HBM and a coordination bus for synchronization is shown in Fig.~\ref{fig:arch_overview}. This bus is responsible for loading instructions into each PU's queue and managing flow control signals for independent or cooperative operations among PUs. The PUs can process multiple DNN layers in parallel and exchange inter-layer activations between different PUs or within the same PU via the shared HBM. This work focuses on the PU architecture and its performance, omitting the description of the instruction controllers and the flow control mechanisms.

\vspace{-0.25cm}
\subsection{Processing Unit}
\begin{figure*}[!t]
\centering
\includegraphics[width=0.975\textwidth]{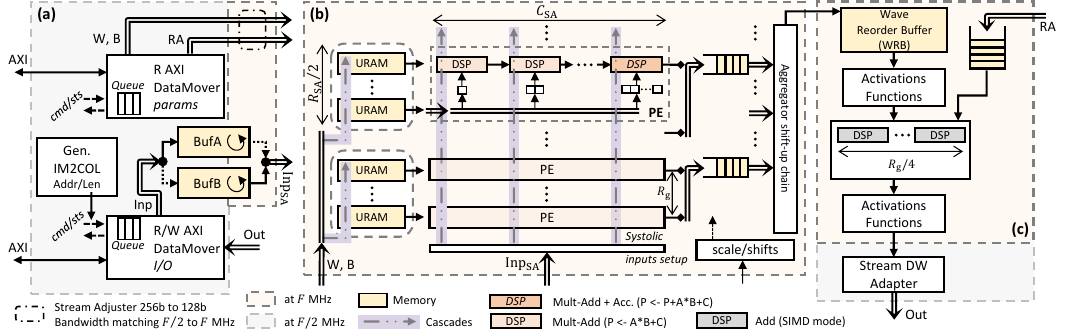}
\vspace{-0.25cm}
\caption{The processing unit (PU) architecture: (a) the pre-processing block, (b) the systolic array, and (c) the post-processing block.}
\label{fig:hw_arch}
\vspace{-0.5cm}
\end{figure*}

\vspace{-0.05cm}
The PU architecture illustrated in Fig.~\ref{fig:hw_arch} consists of three functional blocks primarily outlined as: (a) the pre-processing block containing two AXI DataMover (ADMs) modules interfacing with the HBM, and the Block RAM (BRAM)-based activations buffer, (b) the SA of DSP48E2 units operating with INT8 multiply-accumulate arithmetic alongside a URAM-based weights and biases memory structure, and (c) the post-processing block, i.e., activations functions, residual additions. Two synchronous clock domains are employed in this architecture: a system clock for AXI data transfers and a faster clock (twice as high) for the SA and on-chip memories, with their domains color-coded in Fig.~\ref{fig:hw_arch}. 

% textwidth

A ping-pong buffering scheme (among two BRAMs) is utilized to ensure a continuous stream of activations to the SA, as shown in Fig.~\ref{fig:hw_arch}(a). In essence, while one buffer is being read at the fast clock rate with $C_{\mathrm{SA}}$-elements width to feed the SA, the other is loaded by the ADM \textit{I/O} module using a wider data bus at the system clock rate. Besides regular data transfer support, this ADM command interface is coordinated with an Image to Column (IM2COL) transformation module, allowing a common input datapath to the activations buffers, as presented in the following subsection~\ref{sec:hw_arch_comp}. Moreover, the weights and bias values are allocated in URAM blocks (placed in a column, see Fig.~\ref{fig:hw_arch}(b)), where the write ports exploit the cascades for data, addresses, and control signals \cite{amd_uram_wp}. The same strategy is used in \cite{samajdar2019scaling} for the weights in their Matrix-Vector Multiplication (MVM) tiles. This URAM cascade structure is updated through the ADM \textit{params} module, which transfers the weights to the appropriate locations in each URAM block. Also, it is used to load the bias elements in the spare byte of each URAM block when ECC configuration is not utilized. These bias values remain unaltered at runtime.

Figure~\ref{fig:hw_arch}(b) depicts the SA, which contains $R_{\mathrm{SA}}$ rows and $C_{\mathrm{SA}}$ columns of DSP48E2 instances. In the steady-state pipeline, each row computes a $C_{\mathrm{SA}}$-length dot-product in parallel, denoted as processing element (PE), as all rows share a common activation input, which varies on each column. These inputs enter from the bottom of each column and propagate upwards via the DSP48E2 input cascade paths. To allow PEs to fetch weight matrix rows and bias elements in parallel, each URAM read port operates independently and is enabled in an aligned systolic fashion as the inputs advance through the SA, providing $C_{\mathrm{SA}}$ weight elements in each PE and a bias value in its first column DSP C-port. Subsequently, the partial products flow horizontally, with each DSP passing intermediate dot-product results to its neighbor DSP C-port \cite{amd_dsp48e2_guide} from left to right. In the last column of SA, the partial products are accumulated for $\lceil M / C_{\mathrm{SA}} \rceil$ rounds to complete an MVM with an $R_{\mathrm{SA}} \! \times \! M$ weight matrix.
 
After power-of-two scaling by the scale/shifts module (see Fig.~\ref{fig:hw_arch}(b)), the adjusted systolic wave MVM results are merged and aligned to form a data chunk of $R_{\mathrm{g}}$ bytes for each of the $R_{\mathrm{SA}} / R_{\mathrm{g}}$ row-blocks. This chunk is then pushed to FIFOs that are lane-aligned to the last PE of each row-block. These shallow-depth FIFOs stream data, tagged with the row-block ID and the wave ID, into the aggregator shift-up register chain. In particular, this chain contains multiplexers and registers in each FIFO lane to coordinate up/downstream data transfers. Since the SA produces $R_{\mathrm{SA}}$ bytes with an interval of $\lceil M / C_{\mathrm{SA}} \rceil$ cycles, the aggregator could provide these chunks to the Wave Reorder Buffer (WRB) in an out-of-order context, depending on the MVM and the SA dimensions. Therefore, having each WRB write entry tagged, new waves can be provided to the buffer, thus minimizing the idle state of the pipeline and exhibiting high computational efficiency.

On the contrary, the read side of the WRB enforces strict ordering of the waves, streaming the MVM results to the post-processing modules that direct the results to the non-linear activation functions first (if applicable), i.e., rectified linear unit (ReLU), as shown in Fig.~\ref{fig:hw_arch}(c). An element-wise addition unit is used when the output is fused with a residual path from a prior convolutional layer, as in ResNet architectures, hence avoiding extra off-chip memory transfers \cite{xing2019dnnvm}. For this purpose, we utilized the DSP48E2 SIMD mode and implemented $R_{\mathrm{g}} / 4$ such units. Then, the results are passed again by the required activation function. Finally, the resulting data are adapted with a width upsizing module and a clock conversion unit to match the ADM \textit{I/O} width before being written to HBM.

\vspace{-0.3cm}
\subsection{Computational Dataflow} \label{sec:hw_arch_comp}
\begin{figure*}[!t]
\centering
\includegraphics[width=0.95\textwidth]{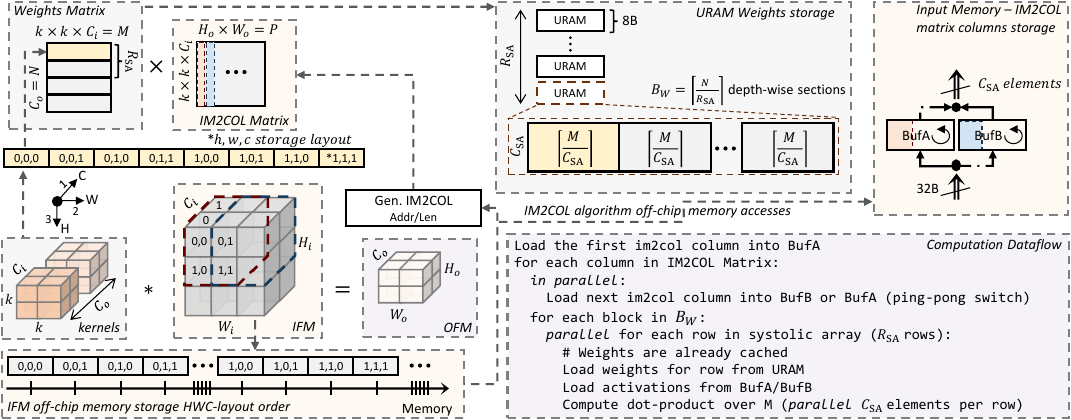}
\vspace{-0.3cm}
\caption{The transformation of a convolutional layer to matrix multiplication and the computational dataflow pseudocode.}
\label{fig:conv_dataflow}
\vspace{-0.5cm}
\end{figure*}

\vspace{-0.05cm}
According to the PU architecture, which supports MVM operations, it is desirable to incorporate a dataflow wrapper that performs General Matrix-Matrix Multiplication (GEMM) operations and transforms Conv layers into GEMM operations. To achieve this, we placed a hardware unit to realize the IM2COL procedure (see Fig.~\ref{fig:hw_arch}(a)), which is a transformation of Conv weights and activations from 4D to 2D format in specific ordering layouts, as shown in Fig.~\ref{fig:conv_dataflow} (left-part). This dedicated unit generates address and length bundles provided to the ADM \textit{I/O} command interface according to the input feature map (IFM) dimensions and Conv characteristics. Hence, the ADM transfers feature-map segments from HBM, which are in height-width-channel (HWC) order, and reshapes them into IM2COL patches, effectively forming the IM2COL matrix. As a result, it supports arbitrary kernel sizes, strides, and padding configurations. The decoupling of the IM2COL technique from the activation buffer, by moving its complexity as ADM commands, simplifies the inter-layer activations management. Specifically, we implemented a common PU input datapath that supports both linear and stride-patterned data transfers for FC and Conv layers with parameters $k\!=\!1,\,p\!=\!0,\,s\!\in\!\{1,2\}$ (kernel size, padding, and stride, respectively), while enabling IM2COL-based transfers for other convolution configurations. This unification allows the PU to operate without benchmark-dependent FPGA reconfiguration, compared to \cite{samajdar2019scaling}, where a Space-Division multiplexing is used to select the ratio between MVM and Conv tiles, according to the DNN model, and also the Conv tile re-implementation for layers with $k\!\neq\!3$.

% textwidth

Assuming GEMM or Conv execution in the PU, for a $N \! \times \! M$ weight matrix and an $M \! \times \! P$ activation matrix, we have a sequence of $P \! \times \! \lceil N / R_{\mathrm{SA}} \rceil$ input rounds to the SA. The weight matrix is stored in $\lceil N / R_{\mathrm{SA}} \rceil$ depth-wise sections in each URAM of $\lceil M / C_{\mathrm{SA}} \rceil$ entries ($C_{\mathrm{SA}}$ elements each). The SA reuses the same input buffer $\lceil N / R_{\mathrm{SA}} \rceil$ times to compute all rows of the weight matrix for an MVM. Meanwhile, the alternate buffer is filled with the next IM2COL column, enabling a seamless pipeline when the loading time is less than the computation time. As shown in Fig.~\ref{fig:conv_dataflow}, this approach is common to Conv and FC layers, which run as GEMMs.

\vspace{-0.25cm}
\section{Weight Transfer Scheduling Strategy}
\vspace{-0.05cm}
We consider the inference of a model in each PU, layer by layer, with every layer $v$ partitioned into weight matrix tiles of size $R_{\mathrm{SA}} \! \times \! M_{v}$. The weights are stored in the HBM region during initialization, and a weight transfer scheduler moves them on-chip while inference is running, intending to minimize tile loading stalls. Since each tile matches the URAM row dimension, only their column entry capacity is tracked.

The $i$-tile is associated with weights load time $\ell_i$ (HBM to URAM), execution time $e_i$ (once loaded), and URAM usage. During inference, the sum of allocated URAM entries cannot exceed its capacity, and once a tile completes its execution, its allocated URAM region can be utilized by next tiles. Tiles for which the weights load time does not exceed the previous tile's execution time ($\ell_i \! \le \! e_{i-1}$) and enough URAM memory is available, exhibit zero pipeline stall. Otherwise, they are identified as stall sources, and the pipeline waits for $\ell_i \! - \! e_{i-1}$, or $\ell_i$ if the memory space is the limiting factor instead.

\begin{figure}[!t]
\centering
\includegraphics[width=1.0\columnwidth]{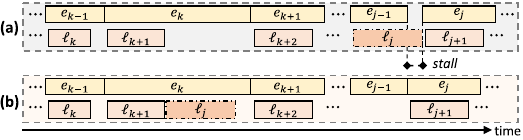}
\vspace{-0.75cm}
\caption{Example of two-phase scheduling: (a) baseline, and (b) adaptive.}
\label{fig:scheduling}
\vspace{-0.5cm}
\end{figure}

A two-phase heuristic (baseline and adaptive) approach is proposed to reduce these stalls, as shown in Fig.~\ref{fig:scheduling}. In the baseline phase, loading the next tile's weights is attempted while the preceding tile operates. Then, the adaptive phase examines potential stalls from the first phase to determine if they can be shifted into earlier processing windows, each corresponding to a tile’s execution time and serving to conceal the weights load time of subsequent tile(s). For each stalled $(j\! - \!1)$-tile (in descending order of stall duration), the method searches for a prior tile $k$ with processing time $e_k$ and adequate memory space to conceal $\ell_j$. If $\ell_j$ can be relocated fully, the $(j\! - \!1)$-tile no longer exhibits a stall. Any relocation that reduces the overall stall time is retained, otherwise it is reversed, and the method proceeds to examine earlier tiles. The same procedure is applied to the next stalled tiles.

% columnwidth

\vspace{-0.25cm}
\section{Implementation}
\vspace{-0.05cm}
The proposed design was prototyped in RTL on an AMD Alveo U50 (XCU50) FPGA by placing multiple PUs, as shown in Fig.~\ref{fig:arch_overview}, without crossing the Super Logic Regions (SLRs) to meet timing constraints. Each PU utilizes one column of maximum URAM blocks ($64$) and a subset of the total DSP48E2 columns ($32$). Since the device provides five URAM columns per SLR, three $\text{PU}_{2\mathrm{x}}$ ($R_{\mathrm{SA}}\!=\!64,\, C_{\mathrm{SA}}\!=\!8$) and two $\text{PU}_{1\mathrm{x}}$ ($R_{\mathrm{SA}}\!=\!64,\, C_{\mathrm{SA}}\!=\!4$) were placed in the upper SLR, with the row-block parameter ($R_{\mathrm{g}}$) set to 8 on both. In contrast, three $\text{PU}_{1\mathrm{x}}$ and two $\text{PU}_{2\mathrm{x}}$ were implemented in the lower SLR to accommodate also a PCIe Gen3 $\! \times 8$ subsystem. In addition, in $\text{PU}_{1\mathrm{x}}$ instances, each URAM is partitioned into two sub-regions, matching the 32-bit ($C_{\mathrm{SA}}$) weight read data path.

Each PU is connected to two 256-bit AXI ports, which interface with the HBM controller operating at 450 MHz and bridging transactions into the 300 MHz AXI domain. One port handles input/output streams, while the other fetches weights, biases, and residual inputs. A stream-width adapter for the latter converts the 256-bit interface at 300 MHz (system clock) to 128 bits at 600 MHz (fast clock), aligned with the URAM data widths, when splitting each PU's URAM weight structure to two components of URAM blocks cascades of $R_{\mathrm{SA}}/{2}$ length each (see Fig.~\ref{fig:hw_arch}(b)). Each PU type uses 20 BRAMs, 64 URAMs, and distinct counts of LUT and DSP48E2 resources: 15.5K and 258 for $\text{PU}_{1\mathrm{x}}$, and 16.5K and 514 for $\text{PU}_{2\mathrm{x}}$, respectively. The design with the maximum number of PUs (5 $\text{PU}_{1\mathrm{x}}$ and 5 $\text{PU}_{2\mathrm{x}}$) occupies 100.0$\%$ of the available URAMs, 64.8$\%$ of DSP48E2s, 25.6$\%$ of BRAMs, and 23.4$\%$ of LUTs. The above resources utilization includes additional logic for the PCIe subsystem, the HBM controller, the AXI interconnects, and the instruction controllers.

\vspace{-0.25cm}
\section{Performance Evaluation} \label{sec:perf_eval}
\begin{figure}[!t]
\centering
\includegraphics[width=1.0\columnwidth]{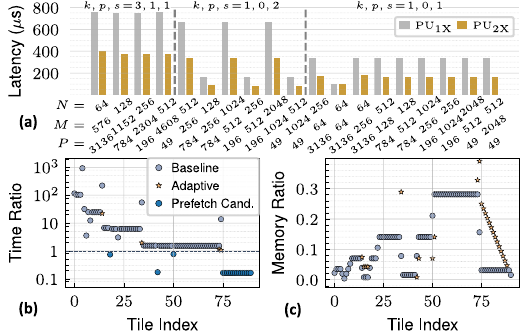}
\vspace{-0.65cm}
\caption{(a) ResNet-50 individual layers latencies for both PU configurations. Two-phase method (b) time and (c) memory ratios for ResNet-18 on $\text{PU}_{2\mathrm{x}}$.}
\label{fig:eval}
\vspace{-0.5cm}
\end{figure}

\vspace{-0.05cm}
We evaluated our design on ResNet-18 and ResNet-50 ImageNet models \cite{He_2016_CVPR}, using 8-bit quantization with power-of-two scaling factors for activations, weights, and biases. Figure~\ref{fig:eval}(a) presents latencies for both PU configurations, measured by hardware counters, for the ResNet-50 convolutional layers. These latencies were measured when the weights of all layers were in the on-chip memory, incurring no tile-loading stalls, and represent the entire pipeline, from fetching activations from HBM to storing outputs back to HBM. Additionally, in these measurements, the WRB buffer read rate ($R_{\mathrm{g}}$ bytes per cycle) exceeds the SA write rate $\bigl(R_{\mathrm{g}} \! \ge \! R_{\mathrm{SA}}/\lceil M / C_{\mathrm{SA}}\rceil\bigr)$, leading to near-optimal per layer throughput efficiency. Conv layers fused with residual additions perform similarly to non-fused ones as the ADM bandwidth sustains weights and residual input transfers of tile execution for our experiments.

To illustrate the efficacy of the two-phase scheduling approach, we report the time and memory ratios for ResNet-18 on $\text{PU}_{2\mathrm{x}}$, as the $\text{PU}_{1\mathrm{x}}$ scheduling is relatively more straightforward due to larger execution times. In the baseline phase, the time ratio compares $i$-tile execution time against $(i\! + \!1)$-tile load time, where a ratio $\! > \!1$ implies complete overlap of load and execution, while a ratio $\! < \!1$ indicates partial overlap. The time ratios are considered in conjunction with the memory ratios, which indicate whether the current and the next tile can fit simultaneously into URAMs. Then, the adaptive phase relocates the remaining stalls into earlier compute intervals. In Fig.~\ref{fig:eval}(b), we show that the blue-colored tiles are successfully relocated to earlier tile indexes, effectively hiding their loading times in the adaptive phase, while Fig.~\ref{fig:eval}(c) confirms that the schedule's memory constraints are satisfied for all tiles. For the initial inference pass, the first tile is pre-loaded to avoid an initial delay, and for subsequent passes, its weights are transferred during the execution of one of the next tiles. Similar ratio measurements were observed for ResNet-50, omitted in this brief due to space limitations. Since the models evaluated here exceed the PU's URAM capacity when the weights are statically allocated, our weight transfer scheduling naturally supports larger models by dynamically allocating weights and reducing their loading stalls. In contrast, prior works \cite{Fan2022Accel, samajdar2019scaling, d2022xdnn} provide only a high-level description of weight transfers without detailing an approach for cases where the weight footprint exceeds on-chip storage.

% columnwidth
In our throughput experiments, each PU independently processes one frame per inference pass, using its own three HBM channel regions, one for weights and two for activations. The activation channels, arranged under the same HBM mini-switch, are statically configured so that residual shortcut activations reside separately from regular activations, eliminating memory congestion and minimizing latency \cite{huang2021shuhai}. The average-pooling layer was executed as a Conv layer, consistent with the approach described in \cite{abdelfattah2018dla}, while the max-pooling layer was not included, as it could be fused into each PU's post-processing block without compromising the achieved rate. Since the IM2COL unit utilizes ADM transfers from HBM, a 32-byte minimum transfer size is required due to alignment constraints. Consequently, we operated the first Conv layer as GEMM by preconfiguring the IFMs on the host into IM2COL matrices and padding each patch to 160 bytes (from 147 bytes to meet alignment) before storing them in the HBM for each PU. This 3-channel layer configuration imposes a modest computational trade-off on inference throughput when using the hardware unit instead, as presented in Table~\ref{tab:fps_comparison}. However, this compromise is acceptable, given that our IM2COL architecture is optimized for high performance in the subsequent layers.

Table~\ref{tab:fps_comparison} compares our accelerator with prior designs for the processing rate in frames per second (FPS) and FPS normalized by Tera Operations per Second (TOPS). Since the implemented DSP48E2-based TOPS of SAs varies across FPGA devices and architectures, the normalized metric reflects each accelerator's efficiency in utilizing the used DSP resources for the benchmark targets. On ResNet-18, our accelerator reaches $0.88\times$ the throughput-optimized Vitis AI DPU's rate \cite{amd_pg367, amd_u50_vitis_ai} but attains a $1.40\times$ improvement in FPS/TOPS. On ResNet-50, our design achieves $1.34\times$ to $1.95\times$ gains in FPS/TOPS compared to the architectures from top to bottom in Table~\ref{tab:fps_comparison}, demonstrating its effective utilization of DSP resources while maintaining high overall throughput. Although the design's RTL and the computational dataflow were not optimized for energy efficiency, we measured the power consumption on ResNet-50 using the on-board sensors. The FPGA consumed on average 46 W (peak 50 W), with 8$\%$ used by the HBM, yielding 12.7 FPS/W ($1.28\times$ higher than \cite{d2022xdnn}) and up to 98$\%$ performance efficiency, defined as the ratio of measured to available TOPS in the SAs. The latter reflects the efficacy of WRB's out-of-order systolic wave support and the weight transfer scheduling in reducing stalls. The accelerator's latency for ResNet-50 is 25.3 and 12.9~$\mathrm{ms}$ for $\text{PU}_{1\mathrm{x}}$ and $\text{PU}_{2\mathrm{x}}$ respectively, and it remains constant for batch sizes 1–5 since it utilizes five instances of each PU type in parallel. 

\begin{table}[!t]
\centering
\caption{Performance Comparison.}
\label{tab:fps_comparison}
\vspace{-0.3cm}
\renewcommand{\arraystretch}{0.95} % Reduces row height to 95% of default
\scriptsize
\begin{tabularx}{\columnwidth}{@{}l *{4}{>{\centering\arraybackslash}X}@{}}
\toprule
Architecture & \multicolumn{2}{c}{FPS} & \multicolumn{2}{c}{FPS/TOPS\textsuperscript{1}} \\
\cmidrule(lr){2-3}\cmidrule(lr){4-5}
Ultrascale+ FPGAs & ResNet-18 & ResNet-50 & ResNet-18 & ResNet-50 \\
\midrule
XCU50 - Ours & 1237.7\textsuperscript{2} & 584.9\textsuperscript{2} & 268.6  & 126.9 \\
XCKU15P - \cite{Fan2022Accel} & - & 242.1 & - & 94.6 \\
XCU50 - \cite{amd_pg367, amd_u50_vitis_ai} & 1410.3 & 572.7 & 191.3 & 77.7 \\
XCVU37P - \cite{samajdar2019scaling} & - & 766.0 & - & 68.8 \\
XCU250 - \cite{d2022xdnn} & - & 1281.0 & - & 65.2 \\
\bottomrule
\end{tabularx}
\vspace{-0.2cm}
\begin{flushleft}
\scriptsize
\textsuperscript{1} Ours: 4.608, \cite{Fan2022Accel}: 2.560, \cite{amd_pg367}: 7.373, \cite{samajdar2019scaling}: 11.140, \cite{d2022xdnn}: 19.660 DSP48E2 TOPS, calculated from the reported numbers of each work, focusing solely on the SAs.\\
\textsuperscript{2} When first conv.\ layer IM2COL in FPGA: ResNet-18: 767.9, ResNet-50: 453.7 FPS.
\end{flushleft}
\vspace{-0.65cm}
\end{table}

\vspace{-0.25cm}
\section{Applicability on an AIMC Emulator}
\vspace{-0.05cm}
In this section, we outline how the proposed FPGA-based accelerator can be extended to emulate the noise characteristics of AIMC devices \cite{LeGallo2023}, enabling system accuracy assessment in DNNs. To emulate AIMC noise, the targeted layers' weights must be embedded with new noise instances at each inference round to capture device-level variations \cite{Petropoulos2020, LeGallo2023_AIHWKIT}. In our architecture, a PU can be substituted with a specialized noise injection unit (NIU) that integrates AIMC noise models, as \cite{LeGallo2023_AIHWKIT}, and uses the same HBM interface as its PU counterpart. The resources required for this module are fewer than those in the PU it replaces, and the data transfer rate is the same to a single HBM channel for reading, modifying, and writing the updated weights. Each PU executes inference on its assigned tiles independently of the activation of AIMC emulation. During inference with noise emulation, the NIU reads noiseless weights of all AIMC tiles from a separate HBM region, injects noise, and overwrites the areas used by the PU. As a result, the PU uses updated noisy weights in each inference round.

In cases where only a subset of weights is dynamically loaded while others remain statically allocated on-chip, an additional mechanism is required to transfer updated noisy weights. However, our weight transfer scheduling fetches weights from HBM in every inference round, ensuring updated weights are inherently incorporated regardless of AIMC emulation. Also, our activations dataflow path to/from HBM allows the storage of output activations for subsequent noise analysis per emulated layer, a capability often unavailable in designs relying solely on on-chip memories for inter-layer activations.

This approach can offer insights for next-generation heterogeneous AIMC chips \cite{Boybat2024}, exploring hybrid mapping schemes, where only specific model layers are AIMC-emulated while others remain on conventional PUs. Additional substitutions can be formed depending on the model and how many emulated PUs are required. In the worst case, pairing an NIU with each PU halves the number of available PUs relative to the original design, whereas if performance requirements allow, a single NIU can be shared by multiple PUs.

\vspace{-0.25cm}
\section{Conclusion}
\vspace{-0.05cm}
This paper presented a configurable FPGA-based accelerator incorporating systolic arrays, HBM, and URAMs to deliver high-throughput DNN inference. One PU configuration leverages higher DSP counts, while the other halves that, using replicated interfaces and peripheral blocks. By instantiating multiple PUs and utilizing a heuristic weight transfer scheduling approach, the accelerator achieves notable gains in throughput and energy efficiency over prior works on ResNet models. Integrating AIMC noise models could seamlessly extend this architecture to a versatile emulation testbed.

\vspace{-0.25cm}
\bibliographystyle{IEEEtran}
\bibliography{IEEEabrv,cites_opt}

\begin{thebibliography}{10}
\providecommand{\url}[1]{#1}
\csname url@samestyle\endcsname
\providecommand{\newblock}{\relax}
\providecommand{\bibinfo}[2]{#2}
\providecommand{\BIBentrySTDinterwordspacing}{\spaceskip=0pt\relax}
\providecommand{\BIBentryALTinterwordstretchfactor}{4}
\providecommand{\BIBentryALTinterwordspacing}{\spaceskip=\fontdimen2\font plus
\BIBentryALTinterwordstretchfactor\fontdimen3\font minus \fontdimen4\font\relax}
\providecommand{\BIBforeignlanguage}[2]{{%
\expandafter\ifx\csname l@#1\endcsname\relax
\typeout{** WARNING: IEEEtran.bst: No hyphenation pattern has been}%
\typeout{** loaded for the language `#1'. Using the pattern for}%
\typeout{** the default language instead.}%
\else
\language=\csname l@#1\endcsname
\fi
#2}}
\providecommand{\BIBdecl}{\relax}
\BIBdecl

\bibitem{yu2019opu}
Y.~Yu, C.~Wu, T.~Zhao, K.~Wang, and L.~He, ``{OPU: An FPGA-Based Overlay Processor for Convolutional Neural Networks},'' \emph{{IEEE} Trans. {VLSI} Syst.}, vol.~28, no.~1, pp. 35--47, Jan. 2020.

\bibitem{abdelfattah2018dla}
M.~S. Abdelfattah \emph{et~al.}, ``{DLA: Compiler and FPGA Overlay for Neural Network Inference Acceleration},'' in \emph{Proc. 28th Int. Conf. Field Program. Log. Appl. (FPL)}, Aug. 2018, pp. 411--4117.

\bibitem{xing2019dnnvm}
Y.~Xing \emph{et~al.}, ``{DNNVM: End-to-End Compiler Leveraging Heterogeneous Optimizations on FPGA-Based CNN Accelerators},'' \emph{{IEEE} Trans. Comput.-Aided Design Integr. Circuits Syst.}, vol.~39, no.~10, pp. 2668--2681, Oct. 2020.

\bibitem{d2022xdnn}
P.~D'Alberto, V.~Wu, A.~Ng, R.~Nimaiyar, E.~Delaye, and A.~Sirasao, ``{xDNN: Inference for Deep Convolutional Neural Networks},'' \emph{ACM Trans. Reconfigurable Technol. Syst.}, vol.~15, no.~2, pp. 1--29, Jan. 2022.

\bibitem{Wu2024Amoeba}
X.~Wu, M.~Wang, J.~Lin, and Z.~Wang, ``{Amoeba: An Efficient and Flexible FPGA-Based Accelerator for Arbitrary-Kernel CNNs},'' \emph{{IEEE} Trans. {VLSI} Syst.}, vol.~32, no.~6, pp. 1086--1099, Jun. 2024.

\bibitem{wang2021autosa}
J.~Wang, L.~Guo, and J.~Cong, ``{AutoSA: A Polyhedral Compiler for High-Performance Systolic Arrays on FPGA},'' in \emph{Proc. ACM/SIGDA Int. Symp. Field-Program. Gate Arrays (FPGA)}, Feb. 2021, pp. 93--104.

\bibitem{li2024revealing}
J.~Li, T.~Li, G.~Shen, D.~Zhao, Q.~Zhang, and Y.~Zeng, ``{Revealing Untapped DSP Optimization Potentials for FPGA-Based Systolic Matrix Engines},'' in \emph{Proc. 34th Int. Conf. Field Program. Log. Appl. (FPL)}, Sep. 2024, pp. 197--203.

\bibitem{samajdar2019scaling}
A.~Samajdar, T.~Garg, T.~Krishna, and N.~Kapre, ``{Scaling the Cascades: Interconnect-Aware FPGA Implementation of Machine Learning Problems},'' in \emph{Proc. 29th Int. Conf. Field Program. Log. Appl. (FPL)}, Sep. 2019, pp. 342--349.

\bibitem{amd_uram_wp}
\BIBentryALTinterwordspacing
{AMD, Inc., Santa Clara, CA, USA}, ``{UltraRAM Breakthrough (WP477)},'' 2016. [Online]. Available: \url{https://docs.amd.com/v/u/en-US/wp477-ultraram}
\BIBentrySTDinterwordspacing

\bibitem{amd_dsp48e2_guide}
\BIBentryALTinterwordspacing
------, ``{UltraScale Architecture DSP Slice (UG579)},'' 2021. [Online]. Available: \url{https://docs.amd.com/v/u/en-US/ug579-ultrascale-dsp}
\BIBentrySTDinterwordspacing

\bibitem{He_2016_CVPR}
K.~He, X.~Zhang, S.~Ren, and J.~Sun, ``{Deep Residual Learning for Image Recognition},'' in \emph{Proc. IEEE Conf. Comput. Vis. Pattern Recognit. (CVPR)}, Jun. 2016, pp. 770--778.

\bibitem{Fan2022Accel}
X.~Fan, G.~Xie, Z.~Huang, W.~Cao, and L.~Wang, ``{Acceleration of Rotated Object Detection on FPGA},'' \emph{{IEEE} Trans. Circuits Syst. {II}}, vol.~69, no.~4, pp. 2296--2300, Apr. 2022.

\bibitem{huang2021shuhai}
H.~Huang \emph{et~al.}, ``{Shuhai: A Tool for Benchmarking High Bandwidth Memory on FPGAs},'' \emph{{IEEE} Trans. Comput.}, vol.~71, no.~5, pp. 1133--1144, May 2022.

\bibitem{amd_pg367}
\BIBentryALTinterwordspacing
{AMD, Inc., Santa Clara, CA, USA}, ``{DPUCAHX8H Performance (PG367)},'' 2024. [Online]. Available: \url{https://docs.amd.com/r/en-US/pg367-dpucahx8h/Performance}
\BIBentrySTDinterwordspacing

\bibitem{amd_u50_vitis_ai}
\BIBentryALTinterwordspacing
------, ``{Vitis AI (UG1354)},'' 2021. [Online]. Available: \url{https://docs.amd.com/r/1.4.1-English/ug1354-xilinx-ai-sdk/Alveo-U50/U50LV-Data-Accelerator-Card}
\BIBentrySTDinterwordspacing

\bibitem{LeGallo2023}
M.~Le~Gallo \emph{et~al.}, ``A 64-core mixed-signal in-memory compute chip based on phase-change memory for deep neural network inference,'' \emph{Nature Electron.}, vol.~6, no.~9, pp. 680--693, Aug. 2023.

\bibitem{Petropoulos2020}
A.~Petropoulos, I.~Boybat, M.~Le~Gallo, E.~Eleftheriou, A.~Sebastian, and T.~Antonakopoulos, ``{Accurate Emulation of Memristive Crossbar Arrays for In-Memory Computing},'' in \emph{Proc. IEEE Int. Symp. Circuits Syst. (ISCAS)}, Oct. 2020, pp. 1--5.

\bibitem{LeGallo2023_AIHWKIT}
M.~Le~Gallo \emph{et~al.}, ``{Using the IBM analog in-memory hardware acceleration kit for neural network training and inference},'' \emph{APL Mach. Learn.}, vol.~1, no.~4, p. 041102, Nov. 2023.

\bibitem{Boybat2024}
I.~Boybat \emph{et~al.}, ``{Heterogeneous Embedded Neural Processing Units Utilizing PCM-Based Analog In-Memory Computing},'' in \emph{{Proc. IEEE Int. Electron Devices Meeting (IEDM)}}, Dec. 2024, pp. 1--4.

\end{thebibliography}

\end{document}